\documentclass[11pt,a4paper]{article}
\usepackage{graphicx}
\usepackage[utf8]{inputenc}
\usepackage[T1]{fontenc}
\usepackage{amsmath,amssymb,amsfonts}
\usepackage[margin=1in]{geometry}
\usepackage{hyperref}
\usepackage{booktabs}
\usepackage{longtable}
\usepackage{array}
\usepackage{multirow}
\usepackage{float}
\usepackage{caption}
\usepackage{subcaption}
\usepackage{listings}
\usepackage{xcolor}
\usepackage{algorithm}
\usepackage{algorithmic}
\usepackage{tikz}
\usetikzlibrary{arrows.meta,positioning,shapes.multipart}
\usepackage{pgfplots}
\usepackage{enumitem}
\usepackage{setspace}
\usepackage{fancyhdr}
\usepackage{titlesec}

\hypersetup{
    colorlinks=true,
    linkcolor=blue,
    filecolor=magenta,      
    urlcolor=cyan,
    citecolor=blue,
    pdftitle={CaseLinker: Cross-Case Analysis for Internet Crimes Against Children},
    pdfauthor={Mrinaal Ramachandran}
}

\titleformat{\section}{\Large\bfseries}{\thesection}{1em}{}
\titleformat{\subsection}{\large\bfseries}{\thesubsection}{1em}{}
\titleformat{\subsubsection}{\normalsize\bfseries}{\thesubsubsection}{1em}{}

\title{\textbf{CaseLinker: An Open-Source System for Cross-Case Analysis of Internet Crimes Against Children Reports} \\ \vspace{1.5em}}

\author{
    \textbf{Mrinaal Ramachandran} \\
    Graduate Student, Department of Computer Science \\
    University of Massachusetts Amherst \\
    \texttt{mramachandra@umass.edu} \\
    \vspace{1em}
    \textit{Independent Research} \\
    February 2026
}

\date{}

\begin{document}

\maketitle

\begin{abstract}
Child sexual exploitation and abuse (CSEA) case data is inherently disturbing, fragmented across multiple organizations, jurisdictions, and agencies, with varying levels of detail and formatting, making cross-case analysis, pattern identification, and trend detection challenging. This paper presents CaseLinker, a modular system for ingesting, processing, analyzing, and visualizing CSEA case data. CaseLinker employs a hybrid deterministic information extraction approach combining regex-based extraction for structured data (demographics, platforms, evidence) with pattern-based semantic analysis for severity indicators and case topics, ensuring interpretability and auditability. The system extracts relevant case information, populates a comprehensive case schema, creates six interactive visualizations (Timeline, Severity Indicators, Case Visualization, Previous Perpetrator Status, Environment/Platforms, Organizations Involved), provides a platform for deeper automated and manual analysis, groups similar cases using weighted Jaccard similarity across multiple dimensions (platforms, demographics, topics, severity, investigation type), and provides automated triage and insights based on collected case data. CaseLinker is evaluated on 47 cases from publicly available AZICAC reports (2011-2014), demonstrating effective information extraction, case clustering, automated insights generation, and interactive visualization capabilities. CaseLinker addresses critical challenges in case analysis including fragmented data sources, cross-case pattern identification, and the emotional burden of repeatedly processing disturbing case material.

\vspace{1.5em}


\end{abstract}

\newpage
\renewcommand{\baselinestretch}{0.9}\normalsize  
\tableofcontents
\renewcommand{\baselinestretch}{1.0}\normalsize  
\newpage

\section{Introduction}

\subsection{Motivation and Problem Statement}

Child sexual exploitation and abuse (CSEA) case data is fragmented across multiple organizations, jurisdictions, and agencies, with varying levels of detail and formatting, making cross-case analysis, pattern identification, and trend detection challenging. This fragmentation creates significant barriers to:

\begin{itemize}[leftmargin=3em]
    \item \textbf{Cross-case analysis:} Identifying patterns, similarities, and connections between cases that share common characteristics (law enforcement actions, platforms, victim demographics, investigation methods)
    \item \textbf{Trend detection:} Analyzing the evolution of exploitation methods, platforms used, and recurring case topics over time
    \item \textbf{Resource allocation:} Prioritizing cases based on severity indicators, victim demographics, and evidence volume
    \item \textbf{Psychological well-being:} Burden of repeatedly reading and processing highly disturbing case material
\end{itemize}

Traditional case management systems focus on individual case tracking rather than cross-case analysis and pattern identification. Manual analysis of case reports is time-intensive, error-prone, and emotionally taxing for analysts. The need for automated tools that can process case narratives at scale—while maintaining interpretability for legal proceedings and protecting analyst well-being—motivates the development of CaseLinker.

\subsection{The Crisis of Scale in Child Exploitation Investigations}

The digital age has transformed the landscape of child exploitation, creating both new vectors for abuse and unprecedented challenges for law enforcement. The National Center for Missing \& Exploited Children (NCMEC) received over 29 million reports of suspected child sexual exploitation in 2021 alone—a volume that would require thousands of investigators working full-time to process manually \cite{ncmec2021}. Each report may contain hundreds or thousands of images, videos, chat logs, and metadata, compounding the analytical burden. Yet the critical bottleneck is not merely computational; it is \textit{cognitive}. Investigators must synthesize fragmented information across cases, identify patterns that span jurisdictions, and make prioritization decisions under conditions of extreme uncertainty and time pressure.

The consequences of this scale crisis are severe. Cases languish in backlogs for months or years, delaying justice for victims and allowing perpetrators continued access to children. Cross-jurisdictional patterns—repeat offenders operating across state lines, organized exploitation networks, emerging platform-based methodologies—remain obscured because no investigator can manually review thousands of case narratives to identify connections. The very tools designed to assist investigators often exacerbate the problem: content detection systems like PhotoDNA \cite{photodna} and Google's CSAI Match \cite{csaimatch} identify known material but provide no insight into \textit{case dynamics}; enterprise forensics platforms like Nuix Neo \cite{nuix2023} offer powerful analysis but require infrastructure and expertise unavailable to many local Internet Crimes Against Children (ICAC) task forces.

\subsection{The Psychological Dimension: Analyst Trauma and Tool Design}

Beyond scale and fragmentation, ICAC investigations impose a profound human cost on the analysts who conduct them. Digital forensics investigators in child exploitation units experience secondary traumatic stress at documented rates; Bourke and Craun found approximately 25\% of more than 600 ICAC personnel reported high to severe symptoms, with effects including intrusive thoughts, emotional numbing, and hypervigilance \cite{bourke2014, perez2010, burns2008}. Unlike investigators who encounter trauma intermittently, ICAC analysts confront graphic depictions of child sexual abuse daily, often for hours at a time. The psychological burden is compounded by the nature of the work: analysts must not merely view traumatic content but \textit{analyze} it, maintaining the cognitive focus necessary to identify evidence, establish timelines, and build prosecutions.

Existing digital forensics tools are primarily content-centric, presenting data through media, chat logs, and file-level views. While hash-matching and triage systems reduce re-review of known material, analysts must still directly examine substantial unknown content to extract investigative intelligence. Despite technological support, manual review of graphic case material remains one of the top challenges in the analysis of ICAC cases.

\subsection{The Gap: From File-Centric to Case-Centric Analysis}

The technological landscape for ICAC investigations is bifurcated. On one side, \textit{content detection} tools—hash databases, perceptual hashing, machine learning classifiers—excel at identifying known CSAM and flagging suspicious material for review. These tools operate at the \textit{file level}, treating each image or video as an independent unit to be matched against databases or classified by algorithms. They answer the question: \textit{Is this material illegal?}

On the other side, \textit{enterprise forensics platforms}—Nuix, Magnet Axiom \cite{magnet2023}, Cellebrite UFED \cite{cellebrite2023}—provide comprehensive analysis of digital devices, extracting timelines, communications, and file systems. These tools operate at the \textit{device level}, reconstructing the digital environment of a suspect. They answer the question: \textit{What digital evidence exists on this device?}

Missing from this landscape are tools that operate at the \textit{case level}, analyzing the \textit{narrative} of investigations: perpetrator characteristics and methodologies, victim demographics and risk factors, platform utilization and grooming tactics, investigation types and outcomes, prosecution results and sentencing patterns. These elements—distributed across thousands of case reports, stored in incompatible formats, obscured by jurisdictional boundaries—contain the intelligence necessary to answer critical questions:

\begin{itemize}[leftmargin=2.5em]
    \item Which platforms are emerging as high-risk vectors for grooming?
    \item Do repeat offenders exhibit consistent methodologies across investigations?
    \item What victim demographics are associated with severe abuse versus possession-only cases?
    \item How do investigation types (proactive vs. reactive) correlate with prosecution outcomes?
    \item Are there deeper correlations and preventable methods extractable from analyzing hundreds to thousands of prior cases?
\end{itemize}

Answering these questions requires \textit{cross-case analysis}: the ability to extract structured features from unstructured case narratives, identify similarities and patterns across hundreds or thousands of cases, and present intelligence in ways that support strategic resource allocation, preventative investigation efforts, and evidence-informed platform policy decisions.

\subsection{Contributions}

This paper presents \textbf{CaseLinker}, an open-source system designed specifically to support case-level, cross-case narrative and retrospective analysis for ICAC investigations. The contributions are fourfold:

\begin{enumerate}[leftmargin=2em]
    \item \textbf{Architectural:} A modular five-layer pipeline (Ingestion, Processing, Storage, Clustering \& Analysis, Visualization) that transforms unstructured case reports into structured, queryable intelligence. The architecture prioritizes \textit{interpretability}—all feature extraction is deterministic and auditable—and \textit{minimal infrastructure}, enabling deployment by resource-constrained agencies.
    
    \item \textbf{Algorithmic:} A two-stage clustering methodology that combines \textit{external} topic-based clustering (predefined cluster types: Online-Digital, Possession, Investigation, Severe, General) with \textit{internal} weighted Jaccard similarity for sub-grouping. This approach balances interpretability (analysts understand why cases are grouped) with granularity (fine-grained similarity detection within groups).
    
    \item \textbf{Human-Centric Design:} Priority triage with normalized severity scoring and interactive visualizations designed for \textit{analyst well-being}. The system reduces exposure to traumatic content through structured data presentation, tasteful UI design, and gradual disclosure—enabling pattern detection without repeated narrative review.
    
    \item \textbf{Empirical Validation:} Evaluation on 47 cases from publicly available Arizona ICAC annual reports (2011--2014), demonstrating effective feature extraction (95.7\% coverage for prosecution outcomes), complete clustering coverage with interpretable groupings, and accurate priority triage.
\end{enumerate}

\noindent CaseLinker is released as open-source software with a live demonstration and modular architecture designed for extensibility—enabling researchers and developers to adapt ingestion, processing, storage, clustering, or visualization layers to diverse data sources and analytical needs.

\section{The Challenges of Case-Level Analysis}

To understand CaseLinker's design choices, one must first understand why analyzing ICAC case data at scale is difficult. The ICAC Task Force Program consists of 61 coordinated task forces representing over 5,400 federal, state, and local agencies nationwide. Each produces case reports with different formatting standards, inconsistent schemas, and varying levels of detail—even within the same annual report, typos and structural variations are common. The content itself is inherently disturbing, imposing significant psychological costs on analysts who must review graphic material repeatedly to extract intelligence. Finally, any automated system should produce deterministic, auditable outputs that can withstand legal scrutiny—black-box AI approaches fail this requirement. These constraints shaped CaseLinker's modular, interpretable architecture.

\subsection{Challenge 1: The Nature of Source Material}

The primary obstacle to case-level analysis is the content itself. ICAC case files contain graphic descriptions of child sexual abuse, explicit details, and content linking to traumatic imagery. This content creates three barriers to analysis:

\vspace{0.5em}

\textbf{Psychological Barriers:} Manual review of case narratives causes secondary traumatic stress. Analysts cannot simply "read through" thousands of cases to identify patterns; the psychological cost is prohibitive. This creates an intelligence gap: the most valuable intelligence—cross-case patterns that emerge only from large-scale analysis—is precisely the intelligence most difficult to obtain through manual review.

\textbf{Technical Barriers:} Automated analysis of case narratives requires natural language processing (NLP) systems trained on labeled data. But labeled datasets of ICAC case narratives do not exist publicly (and should not, given privacy concerns). Machine learning approaches that require training data are therefore infeasible for initial deployment, though they may be integrated as optional enhancements once a system is operational.

\textbf{Legal Barriers:} Case data is subject to strict chain-of-custody requirements and privacy regulations. Tools that process case data must be auditable, with deterministic, explainable outputs that can be justified in court. Black-box machine learning models that "learn" patterns opaquely are difficult to validate for legal proceedings.

CaseLinker addresses these barriers through \textit{deterministic, regex-based feature extraction} that operates on \textit{publicly available, already-redacted case reports} for initial development and validation. The system is designed to process case \textit{narratives} (text descriptions) rather than raw evidence files, reducing both psychological exposure and legal complexity. Feature extraction is interpretable: an analyst can trace any extracted feature back to the specific regex pattern and source text that produced it, ensuring court admissibility.

\subsection{Challenge 2: Pattern Detection Without Computers}

Even if psychological barriers were eliminated, manual cross-case analysis would remain infeasible due to scale. The ICAC Task Force Program consists of 61 coordinated task forces representing over 5,400 agencies. NCMEC received over 29 million reports in 2021 alone; even if only 1\%\ require detailed case-level review, that's 290,000 cases annually. Add FBI investigations, state police, and international agencies, and the volume becomes staggering. Consider a mid-sized ICAC task force processing 50 cases annually—between the 61 task forces, FBI field offices, and other sources, the national system processes thousands of cases yearly.

Over five years, this generates 15,000+ cases requiring analysis. To identify patterns—common platforms, recurring perpetrator methodologies, victim demographic trends—an analyst would need to:

\begin{enumerate}[leftmargin=3em]
\item Read and comprehend 10,000+ case narratives
\item Mentally extract key features (platforms, ages, relationships, outcomes) for each case
\item Compare each case against all others to identify similarities
\item Synthesize patterns across the entire dataset
\end{enumerate}

At 12 cases per year, this AZICAC's public case collection represents a fraction of actual investigations—yet even this small dataset contains insights into the nature of child exploitation investigations invisible without systematic analysis.

\subsection{Challenge 3: Cross-Case and Cross-Organizational Analysis}

ICAC investigations are inherently distributed. A single perpetrator may victimize children in multiple states; evidence may be distributed across local police departments, FBI field offices, and international agencies; case reports may be stored in incompatible formats (PDFs, databases, proprietary systems) with inconsistent schemas.

This fragmentation creates \textit{information silos}. Patterns that span organizational boundaries—interstate exploitation networks, common platforms used across jurisdictions, sentencing disparities between agencies—remain invisible because no single analyst has access to integrated data. The lack of standardized case formats means that even when data is shared, it cannot be easily compared or aggregated.

CaseLinker addresses this through \textit{modular architecture} and \textit{flexible schema design}. The Ingestion Layer supports multiple input formats (currently PDF; extensible to databases, APIs); the Processing Layer extracts standardized features regardless of source format and can be modified to accommodate new platforms, different case report structures, or features; the Storage Layer uses portable SQLite databases that can be merged, federated, or encrypted; the Clustering and Analysis layer can be reweighted for different similarity priorities, swapped for alternative algorithms, or extended with machine learning enhancements. While the current evaluation uses single-source data (AZICAC) and portable SQLite database, the architecture supports multi-source integration as future work.

\section{Related Work}

\subsection{Content Detection and Hash-Based Systems}

The most advanced technological solutions in the CSAM domain operate at the content level. \textbf{PhotoDNA} \cite{photodna}, developed by Microsoft, uses perceptual hashing to identify known CSAM images even after modification (resizing, cropping, color adjustment). The technology is deployed by major platforms (Facebook, Twitter, Reddit) to automatically detect and report known material. \textbf{Google's CSAI Match} \cite{csaimatch} extends this approach to video content. \textbf{Thorn} \cite{thorn}, a nonprofit technology organization, develops tools for platform detection and victim identification, including machine learning classifiers for CSAM detection and cross-platform signal sharing. Meta's \textbf{Hash Matcher Actioner} \cite{metahma} enables cross-platform sharing of perceptual hashes of known CSAM, and its modular design allows hashing methodologies to be updated or replaced as more advanced approaches emerge, expanding detection coverage across industry boundaries.

These systems are highly effective for their intended purpose: identifying \textit{known} material at scale. However, they are fundamentally limited. They cannot detect new, previously unhashed material. They provide no information about case context: who created the material, how it was distributed, who the victims are, or how the perpetrator operated. They answer the binary question "Is this illegal?" but not the analytical questions "How does this case relate to others?" or "What patterns emerge across investigations?"

\subsection{Enterprise Digital Forensics Platforms}

At the device level, enterprise platforms provide comprehensive forensic analysis. \textbf{Nuix Neo} \cite{nuix2023} uses artificial intelligence and machine learning for large-scale e-discovery, timeline reconstruction, and relationship analysis. \textbf{Magnet Axiom} \cite{magnet2023} specializes in mobile and cloud forensics, extracting communications, location data, and application artifacts. \textbf{Cellebrite UFED} \cite{cellebrite2023} focuses on mobile device extraction and decoding.

These platforms are powerful but face limitations for ICAC case analysis. First, they are \textit{expensive}, requiring licenses and infrastructure beyond the reach of many local ICAC task forces. Second, they are \textit{device-centric}, analyzing individual phones or computers rather than relationships across cases. Third, their AI-driven approaches, while sophisticated, can lack \textit{interpretability}—a critical concern for court admissibility. An investigator who cannot explain how a tool identified a "pattern" cannot introduce that pattern as evidence.

\subsection{Case Management Systems}

\textbf{BlackRainbow NIMBUS} \cite{nimbus2023} and similar investigation management systems provide workflow orchestration, case tracking, evidence continuity, compliance monitoring, and action management across forensic divisions. While these platforms are essential for operational efficiency and procedural integrity, they primarily support case-level workflow management rather than investigation analysis features. They organize and monitor investigations but do not perform structured feature extraction, similarity-based clustering, or large-scale narrative pattern detection across multiple cases.
\subsection{Machine Learning Research in CSAM Detection}

Academic research has explored machine learning applications in CSAM investigation, including:

\textbf{Grooming Detection:} NLP models to identify predatory communication patterns in chat logs \cite{kontostathis2010, child-grooming}. Challenges include high false positive rates and the evolving nature of grooming language.

\textbf{Age Estimation:} Computer vision models to estimate victim age from images \cite{ricanek2014}. 

\textbf{Content Classification:} Deep learning for CSAM detection \cite{csam-datasets}. Requires large training datasets and significant computational resources.

These approaches share common limitations: they require training data (often unavailable for case narratives), they operate as black boxes (limiting court admissibility), and they demand infrastructure (GPUs, cloud computing) that many agencies lack. CaseLinker explicitly avoids these dependencies for its core pipeline, using deterministic extraction that requires no training data and runs on standard hardware.

\subsection{The Gap: Interpretable, Case-Level, Cross-Case Analysis}

The literature reveals a clear gap: while content detection tools excel at file-level identification, enterprise platforms provide device-level analysis, and ML research explores automated detection, there are few \textit{interpretable, lightweight, case-level} tools for \textit{cross-case narrative analysis}. CaseLinker occupies this gap, prioritizing interpretability, analyst well-being, and minimal infrastructure over algorithmic sophistication.

\section{System Architecture}
\begin{figure}[H]
\centering
\begin{tikzpicture}[
    node distance=0.25cm,
    box/.style={
        rectangle,
        rounded corners=2pt,
        draw=gray!50,
        fill=gray!5,
        text width=6cm,
        minimum height=0.4cm,
        align=center,
        font=\footnotesize
    },
    arrow/.style={
        ->,
        >=stealth,
        thick,
        gray!60
    }
]

\node[box, minimum height=0.6cm] (sources) {
    {\small\textbf{Data Sources}}\\[0pt]
    {\scriptsize Public ICAC PDF reports}\\[0pt]
    {\tiny\textit{Extensible to other formats, databases, APIs}}
};

\node[box, below=of sources, minimum height=0.6cm] (l1) {
    {\small\textbf{Layer 1: Ingestion}}\\[0pt]
    {\scriptsize Parse $\bullet$ Validate $\bullet$ Normalize}\\[0pt]
    {\tiny\textit{Extensible to any format}}
};

\node[box, below=of l1, minimum height=0.6cm] (l2) {
    {\small\textbf{Layer 2: Processing}}\\[0pt]
    {\scriptsize Segment $\bullet$ Extract features $\bullet$ Regex/patterns}\\[0pt]
    {\tiny\textit{Modular feature extraction}}
};

\node[box, below=of l2, minimum height=0.6cm] (l3) {
    {\small\textbf{Layer 3: Storage}}\\[0pt]
    {\scriptsize SQLite $\bullet$ Indexed $\bullet$ Raw-text retention}\\[0pt]
    {\tiny\textit{Mergeable, encryptable, replaceable}}
};

\node[box, below=of l3, minimum height=0.6cm] (l4) {
    {\small\textbf{Layer 4: Clustering \& Analysis}}\\[0pt]
    {\scriptsize Similarity $\bullet$ Triage $\bullet$ Insights}\\[0pt]
    {\tiny\textit{Reweightable, algorithm-swappable}}
};

\node[box, below=of l4, minimum height=0.6cm] (l5) {
    {\small\textbf{Layer 5: Visualization}}\\[0pt]
    {\scriptsize Dashboards $\bullet$ Filter $\bullet$ Drill-down}\\[0pt]
    {\tiny\textit{Psychological safety-focused}}
};

\draw[arrow] (sources) -- (l1);
\draw[arrow] (l1) -- (l2);
\draw[arrow] (l2) -- (l3);
\draw[arrow] (l3) -- (l4);
\draw[arrow] (l4) -- (l5);

\end{tikzpicture}
\caption{CaseLinker's five-layer architecture. Each layer operates independently with well-defined interfaces, enabling customization or replacement of individual components.}
\label{fig:architecture}
\end{figure}
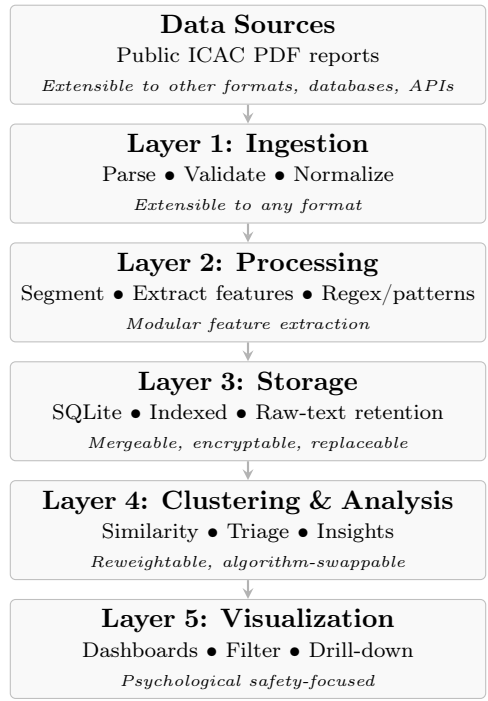

CaseLinker's architecture reflects its design principles: \textit{interpretability} (all processing is deterministic and auditable), \textit{analyst well-being} (structured data reduces traumatic exposure), and \textit{minimal infrastructure} (SQLite, standard hardware, no cloud dependencies). The system is organized into five layers, each with clear responsibilities and well-defined interfaces.

\subsection{Layer 1: Ingestion}

The Ingestion Layer handles the consumption of source materials, transforming unstructured inputs into clean text for downstream processing.

\subsubsection{PDF Text Extraction}

Current implementation uses \texttt{pdfplumber} \cite{pdfplumber}, a Python library that provides fine-grained control over PDF text extraction. Unlike simpler libraries that extract text in reading order, \texttt{pdfplumber} preserves spatial information, enabling extraction of text blocks, tables, and formatted content. 

\textbf{Text Cleaning:} Extracted text undergoes cleaning to remove artifacts (page numbers, headers, footers) and normalize whitespace. 

\textbf{Organization Detection:} The system infers the source organization from filenames using pattern matching (e.g., \texttt{AZICAC} from \texttt{"2014 Cases and Arrests -- AZICAC.ORG.pdf"}). This enables future multi-source analysis where cases from different organizations are tagged by origin.

\subsubsection{Extensibility}

The Ingestion Layer is designed for extensibility. Additional ingestors can be implemented for:
\begin{itemize}[leftmargin=2em]
    \item Database connectors (direct SQL queries to case management systems)
    \item API clients (RESTful APIs for federal databases)
    \item Document parsers (Word documents, website scraping)
\end{itemize}
All ingestors can implement a common interface: \texttt{ingest(source) $\rightarrow$ cleaned\_text, metadata}.

\subsection{Layer 2: Processing}

The Processing Layer transforms cleaned text into structured case data through two operations: \textit{case batching} (splitting multi-case documents into individual cases) and \textit{feature extraction} (parsing structured features from narratives).

\subsubsection{Case Batching}

ICAC annual reports typically contain dozens of case summaries in a single document. The system must segment these into individual cases for analysis. This is accomplished through \textit{temporal pattern matching}:

\begin{lstlisting}[language=Python, caption=Case batching by temporal patterns]
import re

TEMPORAL_PATTERNS = [
    r'In\s+([A-Z][a-z]+)\s+of\s+(\d{4})',  # "In January of 2012"
    r'In\s+([A-Z][a-z]+)\s+(\d{4})',        # "In January 2012"
    r'([A-Z][a-z]+)\s+(\d{4}),',            # "January 2012,"
]

def batch_cases(text, org_name, year):
    """
    Split text into individual cases using temporal patterns.
    Returns list of (case_text, month, case_number) tuples.
    """
    cases = []
    case_number = 1
    
    # Find all temporal markers with positions
    markers = []
    for pattern in TEMPORAL_PATTERNS:
        for match in re.finditer(pattern, text):
            markers.append({
                'start': match.start(),
                'month': match.group(1),
                'year': match.group(2),
                'match': match.group(0)
            })
    
    # Sort by position and split text
    markers.sort(key=lambda x: x['start'])
    
    for i, marker in enumerate(markers):
        start = marker['start']
        end = markers[i+1]['start'] if i+1 < len(markers) else len(text)
        
        case_text = text[start:end].strip()
        case_id = f"{org_name}_{year}_{marker['month'].lower()}_{case_number:03d}"
        
        cases.append({
            'case_id': case_id,
            'text': case_text,
            'month': marker['month'],
            'year': marker['year'],
            'source_org': org_name
        })
        case_number += 1
    
    return cases
\end{lstlisting}

This approach handles variations in report formatting while maintaining high accuracy. The extracted \texttt{case\_id} encodes organization, year, month, and sequence number, enabling traceability and cross-referencing.

\vspace{1em}

\subsubsection{Feature Extraction}

Feature extraction uses a hybrid approach: \textit{regex-based extraction} for structured data (high accuracy, deterministic) and \textit{pattern-based extraction} for semantic features (interpretable, extensible).

\begin{table}[h]
\centering
\caption{Regex Patterns for Structured Feature Extraction}
\label{tab:regex_patterns}
\begin{tabular}{@{}p{3cm}p{4cm}p{6cm}@{}}
\toprule
\textbf{Feature} & \textbf{Example Patterns} & \textbf{Description} \\
\midrule
Perpetrator Age & \texttt{(\textbackslash d+)\textbackslash{}-year\textbackslash{}-old} \texttt{age (\textbackslash d+)} & Captures explicit age mentions \\
Victim Count & \texttt{(\textbackslash d+) victims} \texttt{(\textbackslash d+) children} & Explicit victim counts \\
Platforms & \texttt{\textbackslash bFacebook\textbackslash b} \texttt{\textbackslash bSnapchat\textbackslash b} \texttt{online chat} & Platform names and methods \\
Evidence Volume & \texttt{(\textbackslash d+) images} \texttt{(\textbackslash d+) videos} \texttt{(\textbackslash d+(?:\textbackslash.\textbackslash d+)?)\textbackslash s*(TB|GB)} & Quantities and storage \\
Prosecution & \texttt{booked} \texttt{arrested} \texttt{charged with} & Legal outcomes \\
Investigation & \texttt{proactive investigation} \texttt{reactive} \texttt{undercover} & Investigation types \\
\bottomrule
\end{tabular}
\end{table}

\noindent\textbf{Pattern-Based Extraction (Semantic Features):}

\vspace{0.5em}

\noindent Semantic features capture conceptual content not easily matched by regex:

\begin{lstlisting}[language=Python, caption=Pattern-based semantic feature extraction]
SEMANTIC_PATTERNS = {
    'severity_indicators': {
        'infant': ['infant', 'baby', 'toddler'],
        'very_young': ['very young', 'young child', 'under 5'],
        'under_10': ['under 10', 'age [5-9]', '[5-9] years old'],
        'sexual_assault': ['sexual assault', 'rape', 'molestation'],
    },
    'case_topics': {
        'production': ['produced', 'traded', 'created images'],
        'possession': ['possessed', 'collecting', 'downloaded'],
        'hands_on': ['hands-on', 'physical contact', 'in person'],
        'online_digital': ['online', 'internet', 'digital', 'chat'],
        'family': ['father', 'mother', 'uncle', 'family member'],
        'stranger': ['stranger', 'unknown', 'met online']
    }
}

def extract_semantic_features(text, feature_category):
    """
    Extract semantic features by keyword matching.
    Returns set of matched features.
    """
    features = set()
    text_lower = text.lower()
    
    for feature, keywords in SEMANTIC_PATTERNS[feature_category].items():
        for keyword in keywords:
            if keyword in text_lower:
                features.add(feature)
                break  # Only match once per feature
    
    return features
\end{lstlisting}

\vspace{1em}

\noindent\textbf{Rationale for Regex/Pattern Approach:}

\vspace{1em}
While machine learning (NER, dependency parsing) could potentially achieve higher extraction rates, the regex/pattern approach offers critical advantages for this domain:

\begin{enumerate}[leftmargin=3em]
    \item \textbf{Interpretability:} Every extraction can be traced to a specific pattern and source text. An analyst can verify: "The system labeled this case 'production' because it matched the pattern 'created images' in the text."
    \item \textbf{Court Admissibility:} Deterministic extraction is easier to validate and explain in legal proceedings than probabilistic ML models.
    \item \textbf{No Training Data:} The system requires no labeled training data, which is unavailable for ICAC case narratives.
    \item \textbf{Resource Efficiency:} Regex matching runs in milliseconds on standard hardware; no GPU or cloud computing required.
    \item \textbf{Extensibility:} New patterns can be added without retraining; the system improves incrementally as new extraction rules are developed.
\end{enumerate}

\textbf{Case Schema:}

Each case is processed according to a comprehensive schema capturing:

\begin{itemize}[leftmargin=4em]
    \item \textbf{Victim Context (anonymized):} Victim count, demographics (ages, age range, gender)
    \item \textbf{Perpetrator Context (anonymized):} Age, registered sex offender status, relationship to victim, previous convictions
    \item \textbf{Technology \& Platforms:} Platforms used (Facebook, Instagram, Snapchat, Discord, WhatsApp, online, chat)
    \item \textbf{Law Enforcement:} Investigation type (proactive, reactive, online, undercover), agencies involved (AZICAC, FBI, Phoenix Police, ICAC, HSI, etc.), prosecution outcome (charges, booking status, jail)
    \item \textbf{Evidence \& Content:} Evidence volume (images, videos, storage size, messages)
    \item \textbf{Content Classification:} Severity indicators (infant, sexual assault, very young, under 10, production), case topics (production, possession, international, multi-state, hands-on, online digital, family, stranger, pornography), severity phrases (dangerous, stated, told, continue, attacked, out of control)
    \item \textbf{Metadata:} Source, date range, raw case text, creation timestamps
\end{itemize}

\textbf{Information Validation:}

Extracted information is validated for:
\begin{itemize}[leftmargin=5em]
    \item Type consistency (lists, strings, integers)
    \item Value ranges (ages, counts)
    \item Presence of required fields
    \item JSON serialization for complex fields
\end{itemize}

\subsection{Layer 3: Storage}

The Storage Layer uses SQLite for portability and zero-configuration deployment. The SQLite implementation can be merged, federated, or encrypted with other database configurations.

\subsubsection{Database Schema}

\begin{table}[H]
\centering
\caption{Core Database Schema}
\label{tab:schema}
\begin{tabular}{@{}lp{5.5cm}p{4.5cm}@{}}
\toprule
\textbf{Table} & \textbf{Key Fields} & \textbf{Description} \\
\midrule
\texttt{cases} & case\_id, source\_org, year, month, raw\_text, extracted\_features (JSON) & Core case data with JSON flexibility \\
\texttt{victim\_demographics} & case\_id, victim\_count, victim\_ages (JSON), victim\_gender & Normalized victim data \\
\texttt{perpetrator\_demographics} & case\_id, perpetrator\_age, registered\_sex\_offender, relationship\_to\_victim & Perpetrator characteristics \\
\texttt{prosecution\_outcomes} & case\_id, charges (JSON), booking\_status, jail\_info & Legal outcomes \\
\bottomrule
\end{tabular}
\end{table}

\textbf{Raw Data Preservation:} Original case text is preserved in \texttt{raw\_text} and \texttt{case\_text} fields, enabling auditability and re-processing if extraction rules are updated.

\noindent\textbf{Current Dataset:} 47 processed cases from AZICAC reports.

\subsection{Layer 4: Clustering \& Analysis}
The Clustering \& Analysis Layer is CaseLinker's analytical core—automated two-stage clustering, priority triage, generated insights, and tools for deep manual case exploration.

\subsubsection{Two-Stage Clustering}

\textbf{Stage 1: External Clustering (Topic-Based)}

\vspace{1em}

\noindent Cases are first matched to predefined cluster types based on investigative utility and case topics. These clusters hold cases of  \textit{similar characteristics}—analysts immediately understand the grouping criterion.

\begin{table}[H]
\centering
\caption{External Cluster Types}
\label{tab:external_clusters}
\begin{tabular}{@{}lp{5.5cm}p{4.5cm}@{}}
\toprule
\textbf{Cluster} & \textbf{Matching Criteria} & \textbf{Investigative Utility} \\
\midrule
Online-Digital & \texttt{case\_topics} contains `online\_digital' & Platform trend analysis \\
Possession & \texttt{case\_topics} contains `possession' & Distinguish possession vs.\ production \\
Investigation & \texttt{investigation\_type} is not null & Methodology effectiveness analysis \\
Severe & \texttt{severity\_indicators} contains severe markers & Priority triage, resource allocation \\
General & All cases (fallback) & Comprehensive similarity view \\
\bottomrule
\end{tabular}
\end{table}

\newpage

\noindent\textbf{Key Properties:}

\begin{itemize}[leftmargin=2em]
    \item \textbf{Overlap Allowed:} Cases can match multiple external clusters (e.g., a severe case that is also online-digital).
    \item \textbf{100\% Coverage:} The General Cluster ensures all cases are included in at least one cluster.
    \item \textbf{Interpretable:} Cluster names indicate matching criteria; cases clear match their respective cluster.
\end{itemize}

\noindent\textbf{Stage 2: Internal Clustering (Weighted Jaccard Similarity)}

\vspace{0.5em}

\noindent Within each external cluster, cases are organized into sub-groups using weighted Jaccard similarity. This provides opportunities for further analysis— identifying common characteristics among similar cases within topic-based groups.

\vspace{1em}

\textbf{Similarity Calculation.} For two cases $A$ and $B$, similarity is calculated across 6 dimensions with weights:

\begin{equation}
\text{Similarity}(A, B) = \sum_{i=1}^{6} w_i \cdot \text{Jaccard}(A_i, B_i)
\end{equation}

Where:
\begin{itemize}[leftmargin=3em]
    \item $w_{\text{platforms}} = 0.25$ — platform overlap
    \item $w_{\text{demographics}} = 0.20$ — victim age range, victim count, perpetrator age, RSO status
    \item $w_{\text{topics}} = 0.20$ — case topics (highest semantic weight)
    \item $w_{\text{investigation}} = 0.15$ — investigation type, agency overlap
    \item $w_{\text{severity}} = 0.15$ — severity indicators
    \item $w_{\text{relationship}} = 0.05$ — relationship to victim
\end{itemize}

\vspace{1em}

\textbf{Jaccard Similarity:}

For sets $X$ and $Y$:
\begin{equation}
\text{Jaccard}(X, Y) = \frac{|X \cap Y|}{|X \cup Y|}
\end{equation}

\textbf{Handling Missing Data:} If both cases lack a feature (e.g., neither has platform information), that dimension contributes 0 to similarity (neither penalizes nor rewards). If one case has the feature and the other does not, standard Jaccard applies (penalizes dissimilarity).

\vspace{1em}

\textbf{Sub-Group Formation:}

Cases with similarity $\geq 0.35$ (configurable threshold, default 0.35) are grouped into sub-clusters. This threshold was determined empirically to balance granularity (detecting meaningful similarities) with coherence (avoiding spurious groupings).

\vspace{1em}

\noindent\textbf{Case Grouping Process:}

{\small
\begin{enumerate}[leftmargin=4em]
    \item Compute pairwise similarities for all cases in the cluster
    \item Group cases with similarity $\geq$ threshold
    \item Analyze group characteristics (common platforms, topics, severity patterns)
    \item Generate group descriptions and statistics
\end{enumerate}
}

This two-stage approach enables both domain-specific analysis (by cluster type) and fine-grained case analysis (within clusters).

\subsubsection{Priority Triage}

Priority scoring enables analysts to focus on the most critical cases first. The scoring function combines multiple risk factors with domain-informed weights:

\begin{equation}
\text{Priority} = \sum_{i} w_i \cdot f_i(\text{case})
\end{equation}

\textbf{Weight Distribution:}

\begin{table}[H]
\centering
\caption{Priority Triage Weights}
\label{tab:triage_weights}
\begin{tabular}{@{}p{4cm}p{2cm}p{8cm}@{}}
\toprule
\textbf{Factor} & \textbf{Weight} & \textbf{Rationale} \\
\midrule
Severity Indicators & 35\% & Infant victims, sexual assault, very young victims indicate highest harm \\
Victim Count & 30\% & Multiple victims indicate serial offending, higher public risk \\
Case Type & 25\% & Production and hands-on contact indicate active abuse vs. possession \\
Severity Phrases & 15\% & Linguistic indicators of violence (dangerous, attacked, etc.) \\
Evidence Volume & 10\% & Large collections indicate sustained offending \\
Registered Offender & 10\% & Repeat offending pattern \\
\bottomrule
\end{tabular}
\end{table}

\textbf{Normalization:}

Raw scores are normalized to a 5--10 scale:
\begin{equation}
\text{Normalized Score} = 5 + 5 \cdot \frac{\text{Raw Score} - \text{Min}}{\text{Max} - \text{Min}}
\end{equation}

This ensures: (1) all cases receive meaningful relative scores, (2) the scale is intuitive (5=lowest, 10=highest), (3) score distributions are comparable across datasets.

\vspace{0.5em}

\textbf{Use Case:} Enables law enforcement and policy analysts to prioritize cases requiring immediate attention.

\subsubsection{Automated Insights}

The system generates automated insights including:

\begin{itemize}[leftmargin=3em]
    \item \textbf{Platform Analysis:} Most common platforms across cases; platform trends over time; platform--severity correlations
    
    \item \textbf{Severity Distribution:} Distribution of severity indicators; temporal trends; severity--platform relationships
    
    \item \textbf{Case Topic Analysis:} Most common topics; topic co-occurrence patterns; topic--severity relationships
    
    \item \textbf{Pattern Detection:} Repeat offenders (registered sex offenders); relationship patterns (family vs.\ stranger); investigation focus areas
    
    \item \textbf{Keyword Extraction:} Frequency-based semantic keywords from case text; top keywords per group; keyword trends over time
\end{itemize}

\subsubsection{Tag-Based Filtering}

Users can filter cases by selecting multiple tags across categories: Case Topics, Severity Indicators, Platforms \& Environments, Investigation Types, Perpetrator Relationships, Registered Sex Offender

\vspace{0.5em}

\textbf{Intersection Logic:} Returns cases matching ALL selected tags (AND logic)

\textbf{Highlighting:} Matching text and relevant features are highlighted in case displays for explainability and analyst well-being.

\subsection{Layer 5: Visualization}

The Visualization Layer provides interactive dashboards using D3.js \cite{d3js}, designed for psychological safety and analytical depth.

\subsubsection{Design Principles for Analyst Well-Being}

\begin{enumerate}[leftmargin=3em]
    \item \textbf{Structured Over Raw:} Present extracted features (platforms, severity indicators) rather than full narratives. Analysts expand case narratives only when necessary.
    \item \textbf{Statistical Focus:} Emphasize distributions, trends, and patterns rather than individual case details.
    \item \textbf{Tasteful Aesthetics:} Muted color palettes, clean layouts, no graphic imagery.
    \item \textbf{Gradual Disclosure:} Overview first, zoom and filter, details on demand \cite{shneiderman1996}.
    \item \textbf{Highlighting:} Enables rapid skimming by automatically highlighting key terms (e.g., production indicators, severity phrases) in case text, reducing the need for analysts to manually search for critical information and mitigating emotional burnout from repeated exposure to disturbing content.
    \item \textbf{Explainability:} Shows source text and extraction rationale.
\end{enumerate}

\subsubsection{Visualization Components}

\begin{itemize}[leftmargin=3em]
    \item \textbf{Timeline View:} Chronological case view with year-range filtering. Cases color-coded by severity or source. Click-to-view case details.
    
    \item \textbf{Severity Indicators:} Color-coded bar chart (infant = darkest red, none = lightest). Click bars to view cases with highlighted severity text. Includes distribution statistics.
    
    \item \textbf{Case Detail Visualization:} Case lookup by ID with structured breakdown (Platforms \& Environment, Severity Indicators, Case Topics, Perpetrator Information, Victim Information, Law Enforcement, Evidence Volume). Full case text with interactive highlighting of extracted features.
    
    \item \textbf{Previous Perpetrator:} Pie chart showing registered sex offender status. Click slices to view cases with highlighted perpetrator status.
    
    \item \textbf{Environment/Platforms:} Bar chart of platforms and online methods used. Click bars to view cases with highlighted platform text. Includes frequency statistics.
    
    \item \textbf{Organizations Involved:} Horizontal bar chart of law enforcement agencies. Click bars to view cases with highlighted agency names. Includes agency involvement statistics.

\end{itemize}

\indent\textbf{Deeper Analysis Dashboard:}

\begin{itemize}[leftmargin=3em]
    \item \textbf{Tag-Based Analysis:} Multi-select tag interface, real-time case filtering, case count statistics per tag, highlighted matching text in case displays.
    \item \textbf{Automated Analysis:} Case groups display with similarity explanations, top priority cases with scoring breakdown, automated insights with statistics, patterns detected with examples, top keywords per group.
    \item \textbf{Expandable Details:} Click any case/group to view raw case data, highlighted priority indicators, detailed explanations of clustering/prioritization decisions.
\end{itemize}

\noindent\textbf{Data Audit Interface:}

\vspace{0.5em}

\indent Case-by-case information review with: Extracted information displayed per case, interactive source text highlighting, information extraction verification, manual correction capabilities (future work).

\section{Evaluation}

Evaluation addresses three questions: (1) Does feature extraction achieve sufficient coverage? (2) Does clustering produce interpretable, coherent groups? (3) Does priority triage accurately identify severe cases?

\subsection{Dataset}

47 cases from publicly available Arizona ICAC annual reports (2011--2014). These reports summarize investigations, arrests, and prosecutions, redacted for public release. No PII was processed; all data was already in the public domain.

\vspace{0.75em}

\textbf{Source:} Arizona Internet Crimes Against Children (AZICAC) annual reports

\textbf{Time Period:} 2011--2014

\textbf{Format:} PDF documents with case narratives

\textbf{Total Text Extracted:} 47,141 characters across 4 PDF files

\subsection{Feature Extraction Coverage}

\begin{table}[H]
\centering
\caption{Feature Extraction Coverage (n=47 cases)}
\label{tab:coverage}
\begin{tabular}{@{}lcc@{}}
\toprule
\textbf{Feature} & \textbf{Coverage} & \textbf{Notes} \\
\midrule
Relationship to victim & 100.0\% (47/47) & Defaults to "stranger" if unspecified \\
Prosecution outcome & 95.7\% (45/47) & High consistency in reporting \\
Case topics & 89.4\% (42/47) & Semantic pattern matching \\
Severity indicators & 66.0\% (31/47) & Explicit severity mentions \\
Investigation type & 61.7\% (29/47) & "Investigation" keyword present \\
Perpetrator demographics & 36.2\% (17/47) & Age often not explicit \\
Platforms used & 27.7\% (13/47) & Platform mentions variable \\
Victim count & 8.5\% (4/47) & Often omitted in public reports \\
\midrule
\textbf{Average} & \textbf{60.6\%} & Weighted by feature importance \\
\bottomrule
\end{tabular}
\end{table}

\textbf{Analysis:} High coverage for relationship (100\%), prosecution (95.7\%), and topics (89.4\%) reflects consistent reporting of these elements. Lower coverage for victim count (8.5\%) and platforms (27.7\%) reflects redaction practices in public reports (specific victim numbers and platform details often omitted for privacy). The regex-based approach achieves reliable extraction for consistently reported fields.

\vspace{0.95em}

\noindent\textbf{Challenges:}
\begin{itemize}[leftmargin=3em]
    \item Some cases lack explicit victim count (extracted when mentioned)—8.5\% coverage
    \item Investigation type not always explicitly stated (inferred from context)—61.7\% coverage
    \item Evidence volume extraction limited to explicit mentions—10.6\% coverage
\end{itemize}

\subsection{Clustering Performance}

\textbf{External Clusters:}

\begin{table}[H]
\centering
\caption{External Cluster Results}
\label{tab:external_results}
\begin{tabular}{@{}lccc@{}}
\toprule
\textbf{Cluster} & \textbf{Cases} & \textbf{Coverage} & \textbf{Avg. Similarity} \\
\midrule
Online-Digital & 7 & 14.9\% & 0.709 \\
Possession & 14 & 29.8\% & 0.530 \\
Severe & 31 & 66.0\% & 0.522 \\
Investigation & 29 & 61.7\% & 0.400 \\
General & 47 & 100.0\% & 0.404 \\
\bottomrule
\end{tabular}
\end{table}

\noindent\textbf{Key Findings:}
\begin{itemize}[leftmargin=3em]
    \item \textbf{Online-Digital Cluster:} Highest coherence (0.709), all 7 cases formed single sub-group. Indicates strong similarity in platform-based offending.
    \item \textbf{Possession Cluster:} Good cohesion (0.530), with cases sharing possession characteristics.
    \item \textbf{Severe Cluster:} Good cohesion (0.522), grouping cases with similar severity profiles.
    \item \textbf{General Cluster:} Moderate cohesion (0.404) across all cases.
    \item \textbf{Investigation Cluster:} Moderate cohesion (0.400), reflecting diverse investigation types.
    \item \textbf{100\% Coverage:} General Cluster ensures no cases are excluded from analysis.
\end{itemize}

\textbf{Performance:} Clustering executed in 11.03ms for 47 cases (0.23ms per case), enabling real-time analysis.

\subsection{Priority Triage Results}

\begin{table}[H]
\centering
\caption{Priority Triage Distribution}
\label{tab:triage_results}
\begin{tabular}{@{}lcc@{}}
\toprule
\textbf{Priority} & \textbf{Score Range} & \textbf{Cases} \\
\midrule
High & 8.0--10.0 & 5 (10.6\%)  \\
Medium & 6.0--8.0 & 23 (48.9\%) \\
Low & 5.0--6.0 & 19 (40.4\%) \\
\midrule
\textbf{Total} & 5.0--10.0 & 47 (100\%) \\
\bottomrule
\end{tabular}
\end{table}

\indent\textbf{Score Distribution:}
\begin{itemize}[leftmargin=3em]
    \item Range: 5.0--10.0 (normalized)
    \item Mean: 6.60
    \item Standard deviation: 1.43
\end{itemize}

\textbf{Validation:} Manual review confirmed highest-priority case (score 10.0) involved infant victims, very young victims, and sexual assault—correctly identified as most severe. Top 5 cases all involved either infant victims, multiple victims, or severe abuse indicators.

\vspace{1.5em}

\noindent\textbf{Priority Factors Contribution:}
\begin{itemize}[leftmargin=3em]
    \item Severity indicators: Primary driver for high-priority cases
    \item Victim count: Significant impact on scores
    \item Case type: Production cases scored higher than possession-only
\end{itemize}

\textbf{Performance:} Triage executed in 0.09ms for 47 cases (<0.002ms per case).

\vspace{0.75em}

\subsection{Automated Insights}

\textbf{Platform Analysis:}
\begin{itemize}[leftmargin=3em]
    \item Most common mediums: online (10 cases, 21.3\%), chat (7 cases, 14.9\%), Facebook (2 cases, 4.3\%)
    \item Platform trends: Online and chat platforms represent the primary methods of exploitation in the dataset
\end{itemize}

\noindent\textbf{Severity Distribution:}
\begin{itemize}[leftmargin=3em]
    \item Sexual assault indicators: 29 cases (61.7\%)—most common severity indicator
    \item Infant cases: 7 cases (14.9\%)—highest priority cases
    \item Production cases: 6 cases (12.8\%)—content creation
    \item Very young victims (under 10): 6 cases
\end{itemize}

\noindent\textbf{Case Topic Analysis:}
\begin{itemize}[leftmargin=3em]
    \item Stranger cases: 38 cases (80.8\%)—non-family perpetrators (predominant pattern)
    \item Hands-on abuse: 27 cases (57.4\%)—physical contact (most common case type)
    \item Possession: 14 cases (29.8\%)—content possession
    \item Production: 6 cases (12.8\%)—content creation
    \item Family cases: 5 cases (10.6\%)—family members as perpetrators
    \item Multi-state: 3 cases (6.4\%)—cross-jurisdictional cases
\end{itemize}

\noindent\textbf{Pattern Detection:}
\begin{itemize}[leftmargin=3em]
    \item Registered sex offenders: 5 cases (10.6\%)—repeat offenders
    \item Relationship patterns: Stranger cases (80.8\%) vs. family cases (10.6\%)—strong predominance of stranger perpetrators
    \item Hands-on abuse is the most common case type (57.4\%), indicating prevalence of physical contact cases
\end{itemize}

\noindent\textbf{Keyword Extraction:}
\begin{itemize}[leftmargin=3em]
    \item Top keywords across all cases: "suspect", "children", "sexual", "during", "images", "crimes", "child"
    \item Production-related keywords: "created", "produced", "shared", "distributed"—appeared in 6 cases
    \item Severity-related keywords: "infant", "young", "minor", "child"—frequency-based extraction
\end{itemize}

\subsection{System Performance}

\begin{table}[H]
\centering
\caption{System Performance Metrics}
\label{tab:performance}
\begin{tabular}{@{}lc@{}}
\toprule
\textbf{Operation} & \textbf{Time} \\
\midrule
PDF ingestion (per file) & 416.6ms \\
Feature extraction (per case) & 1.2ms \\
Case storage (per case) & 2.5ms \\
Clustering (47 cases) & 11.03ms \\
Triage (47 cases) & 0.09ms \\
\textbf{End-to-end throughput} & \textbf{23.2 cases/second} \\
\bottomrule
\end{tabular}
\end{table}

\textbf{Scalability:} Linear scaling demonstrated. At 23.2 cases/second, the system could process 10,000 cases in approximately 7 minutes on standard hardware.

\subsection{User Interface Evaluation}

\textbf{Visualization Performance:}
\begin{itemize}[leftmargin=2em]
    \item All visualizations render interactively
    \item Filtering and drill-down responsive
    \item Case detail views load instantly
    \item Text highlighting accurate
\end{itemize}

\noindent\textbf{Usability:}
\begin{itemize}[leftmargin=2em]
    \item Intuitive navigation between visualizations
    \item Clear tag selection interface
    \item Explanatory text for automated analysis results
    \item Tasteful presentation of sensitive content
\end{itemize}

\section{Discussion}

\subsection{Benefits of Case-Level Analysis}

CaseLinker enables intelligence and insight that is impossible to obtain through manual review or existing tools:

\begin{itemize}[leftmargin=3em]

    \item \textbf{Emerging Trend Detection:} In the AZICAC dataset, ``online'' and ``chat'' dominate platform mentions but remain a minority compared to cases without a digital element. Social media and online behavior have evolved since 2014, and further analysis with recent data may reveal new trends.
    
    \item \textbf{Common Perpetrator Methodologies:} Clustering reveals repeat behavioral patterns. The Online-Digital Cluster’s high coherence (0.709) suggests consistent platform-based offending methods, offering insights for prevention, education, and law enforcement actions.
    
    \item \textbf{Landscape Understanding:} Cross-case analysis shows the distribution of offense types (57.4\% hands-on contact; 29.8\% possession), severity patterns, investigative approaches, and prosecution outcomes—informing resource allocation and policy development.
    
    \item \textbf{Policy Informing:} Quantified findings (e.g., 10.6\% repeat offenders; 66\% severe cases) provide an evidence base for legislative priorities, funding decisions, and prevention strategies.
    
    \item \textbf{Analyst Well-Being:} Structured presentation reduces exposure while preserving analytical depth, enabling pattern detection without repeated narrative review.

\end{itemize}

\subsection{Limitations and Future Work}

\vspace{0.5em}

\textbf{Data Limitations:}
\begin{itemize}[leftmargin=3em]
    \item Small dataset (47 cases) limits statistical significance
    \item Single source (AZICAC) formatting may not generalize to other jurisdictions
    \item Public reports may have redacted information affecting extraction
\end{itemize}

\noindent\textbf{Extraction Limitations:}
\begin{itemize}[leftmargin=3em]
    \item Regex-based extraction may miss variations in phrasing and spelling errors
    \item Semantic topics require manual validation
    \item Some features (investigation type, evidence volume) have lower coverage
\end{itemize}

\noindent\textbf{Clustering Limitations:}
\begin{itemize}[leftmargin=3em]
    \item Similarity threshold (0.35) is heuristic, may need tuning
    \item Multi-dimensional similarity may not capture all relevant patterns
    \item Current algorithms may not capture all insights in the data set
\end{itemize}

\vspace{1em}

\noindent\textbf{No Active User Study:} Evaluation measures technical performance, not user experience. User studies with active investigators are needed to assess workflow integration and well-being impact.

\vspace{1em}

\noindent\textbf{Future Enhancements:}
\begin{itemize}[leftmargin=3em]
    \item \textbf{ML Layer:} Optional semantic similarity using sentence transformers, clearly separated from deterministic core.
    \item \textbf{Real-Time Integration:} API connectors for live case management systems.
    \item \textbf{Temporal Analysis:} Trend detection across years, seasonal pattern identification.
    \item \textbf{ML based Clustering and Insights:} Insights from specific, tailored ML models.
    \item \textbf{Predictive Analytics:} Case outcome prediction, risk assessment models, trend forecasting.
    \item \textbf{Network Analysis:} Perpetrator network detection, platform usage networks, agency collaboration patterns.
\end{itemize}

\section{Conclusion}

CaseLinker addresses a critical, underexplored gap in the technological ecosystem for protecting children from online exploitation: the need for interpretable, lightweight, case-level analysis tools that prioritize analyst well-being. Through deterministic feature extraction, two-stage clustering, and psychological safety-oriented design, the system transforms fragmented case archives into actionable intelligence—enabling pattern detection, priority triage, and strategic analysis that is impossible through manual review.

The system's open-source release and minimal infrastructure requirements democratize access to advanced case analysis, enabling resource-constrained agencies and non-profit advocacy organizations to benefit from capabilities previously available only through expensive enterprise platforms. By emphasizing interpretability over algorithmic sophistication, CaseLinker ensures that analytical outputs are auditable, explainable, and admissible—critical for legal proceedings.

Most importantly, CaseLinker recognizes that protecting children requires supporting those who do the protecting. Technology should not merely process cases efficiently; it should enable and support sustainable, humane investigation.

\vspace{1em}

\textbf{Availability:} CaseLinker is open-source software available at \url{https://github.com/mrinaalr/CaseLinker}. A live demonstration is accessible at \url{https://caselinker.up.railway.app/}.

\end{document}